\documentclass[sigplan,10pt,nonacm]{acmart}
\pdfoutput=1
\startPage{1}
\setcopyright{none}
\usepackage{amsmath,amsfonts,amsthm} %
\usepackage{comment}                         %
\usepackage{ifthen}                          %
\usepackage{minibox}                         %
\usepackage{multirow}                        %
\usepackage{suffix}                          %
\usepackage{graphicx}                        %
\usepackage{xargs}                           %
\usepackage[capitalise,noabbrev,nameinlink]{cleveref}            %
\usepackage{hyphenat}                        %
\usepackage{subcaption}                      %
\usepackage{fancyvrb}                        %
\usepackage{xcolor}                          %
\usepackage{supertabular}                    %
\usepackage{xspace}                          %
\usepackage{booktabs}   %
\newcommand\ie{i.e.,\xspace}
\newcommand\eg{e.g.,\xspace}

\newcommand{\TODO}[1]{{\color{red}#1}}

\newcommand{\mypara}[1]{\smallskip\noindent\emph{\textbf{{#1.}}}}

\usepackage{wasysym}
\usepackage{stmaryrd}
\usepackage{mathtools}
\usepackage{mathpartir}
\usepackage[excludeor]{everyhook}
\usepackage{halloweenmath}
\usepackage[normalem]{ulem}

\newtheorem{definition}{Definition}[section]

\newcommand{\qstep}{\rightarrow}
\newcommand{\zstep}{\rightsquigarrow}

\newcommand{\lang}{ZFI{\small$\xrightswishingghost{}$}\xspace}

\newcommand{\ctx}{\Psi}
\newcommand{\pc}{\mathit{pc}}
\renewcommand{\P}{\mathit{P}}
\newcommand{\Reg}{\mathit{Reg}}
\newcommand{\Mem}{\mathit{Mem}}
\newcommand{\ustate}{\mu{}\mathit{state}}

\newcommand{\Obs}{\mathit{Obs}}
\newcommand{\flag}{\mathit{mispredicted}}

\newcommand{\insn}{\mathit{insn}}
\newcommand{\rbase}{r_{\mathit{base}}}
\newcommand{\rheap}{r_{\mathit{Heap}}}
\newcommand{\rstk}{r_{\mathit{Stk}}}
\newcommand{\vstk}{v_{\mathit{Stk}}}
\newcommand{\rsstk}{r_{\mathit{SStk}}}
\newcommand{\vsstk}{v_{\mathit{SStk}}}
\newcommand{\vaddr}{v_{\mathit{addr}}}
\newcommand{\eoff}{e_{\mathit{off}}}

\newcommand{\mgen}[2]{{}^*({#1} + {#2})}
\newcommand{\m}{\mgen{\rbase}{\eoff}}
\newcommand{\incr}{{+\!+}}
\newcommand{\eval}[2][\ctx]{\llbracket{#2}\rrbracket_{#1}}
\newcommand{\ictx}[2][\ctx]{{#1}[{#2}]}
\newcommand{\ctxincr}[1][\ctx]{{#1}^\incr}
\newcommand{\ctxplus}[2][\ctx]{{#1}^\incr \left\{#2\right\}}
\newcommand{\ctxnext}[2][\ctx]{{#1} \{#2\}}

\newcommand{\nothing}{-}

\newcommand{\jmp}{\mathit{jmp}}
\newcommand{\call}{\mathit{call}}
\newcommand{\ret}{\mathit{ret}}

\newcommand{\flush}{\textit{flush}}
\newcommand{\cetendbranch}{\mathit{endbranch}}

\newcommand{\base}{\mathit{base}}

\renewcommand{\cref}[1]{\Cref{#1}}

\newcommand{\mg}[1]{\textcolor{blue}{MG: #1}}

\newcommand{\mv}[1]{\textcolor{olive}{MV: #1}}

\begin{document}

\title{A Turning Point for Verified Spectre Sandboxing}

\author{Sunjay Cauligi}
\affiliation{
  \institution{UC San Diego, USA}
}
\author{Marco Guarnieri}
\affiliation{
  \institution{IMDEA Software Institute, Spain}
}
\author{Daniel Moghimi}
\affiliation{
  \institution{UC San Diego, USA}
}
\author{Deian Stefan}
\affiliation{
  \institution{UC San Diego, USA}
}
\author{Marco Vassena}
\affiliation{
  \institution{CISPA Helmholtz Center for Information Security, Germany}
}

\begin{abstract}

  Spectre attacks enable an attacker to access restricted data in an application's memory.
  Both the academic community and industry veterans have developed several
  mitigations to block Spectre attacks, but to date, very few have been
  formally vetted; most are ``best effort'' strategies.
  Formal guarantees are particularly crucial for protecting isolated
  environments like sandboxing against Spectre attacks.
  In such environments, a subtle flaw in the mitigation would allow untrusted
  code to break out of the sandbox and access trusted memory regions.

  In our work, we develop principled foundations to build isolated
  environments resistant against Spectre attacks.
  We propose a formal framework for reasoning about sandbox execution
  and Spectre attacks. We formalize properties that sound mitigation
  strategies must fulfill
  and we
  show how various existing mitigations satisfy (or fail to satisfy!) these properties.
\end{abstract}

\maketitle

\input{rref}

\section{Introduction}
Software-based Fault Isolation (SFI) is a popular technique for efficiently confining untrusted code to a
\emph{software sandbox}~\cite{Tan:2017}.
For example, web browsers and cloud providers rely on SFI-based
sandboxes to prevent buggy or malicious code from corrupting the
memory of the host and other sandboxes~\cite{lucet-talk,nacl,fastly-wasm}.
Unfortunately, untrusted code can leverage speculative execution to
break out of the sandbox and access trusted memory regions, thus
making existing SFI implementations vulnerable to Spectre attacks~\cite{Kocher2019spectre,project0-spectre-poc}.

Researchers have proposed different approaches to mitigate Spectre
attacks in SFI-style sandboxes~\cite{swivel,shen2019restricting,Jenkins20}.
However, these are best-effort proposals: They rely on carefully combining
several intricate software protections and hardware extensions to prevent
unsafe speculative behaviors.
It is unclear whether the combination of these countermeasures
work as intended and so, in practice, these approaches may fail to
provide the expected security guarantees against Spectre attacks.
This gap is apparent in the security guarantees
of \emph{Swivel}~\cite{swivel}, a WebAssembly sandboxing system---although
Swivel indeed prevents many Spectre attacks, its security guarantees
are conditioned on several underlying assumptions and constraints.
As one example, the \emph{Swivel-CET} implementation prevents speculative leakage
via the data cache, but does not stop leakage via control flow.

In this paper, we develop principled foundations to build reliable
sandboxing mechanisms against Spectre attacks.
Towards this goal, we formulate security properties to
formally capture the essence of Spectre SFI attacks.
We apply our formal framework to investigate Swivel's security claims and 
show which Spectre attacks it can soundly mitigate and for which it falls short.

\section{Background}
\label{sec:background}

We begin with an overview of speculative attacks on SFI sandbox systems
and a brief description of the mitigation techniques that Swivel
employs to enforce speculative sandbox isolation and control-flow integrity.

\subsection{Speculative SFI attacks}
\label{sec:sfi-attacks-background}

Swivel identifies two distinct classes of speculative attacks
on SFI sandboxes: \emph{Breakout attacks} and \emph{poisoning attacks}~\cite{swivel}.
First, the sandbox host system does not trust the individual sandboxes:
Swivel prevents \emph{breakout attacks}, where a sandbox accesses data outside of its
defined memory regions.
Second, Swivel's sandboxes themselves are \emph{mutually distrusting}: Swivel
prevents \emph{poisoning attacks}, where an attacker
is able to leak secrets from a victim sandbox.

\mypara{Breakout attacks}
A sandbox breakout occurs when a
malicious sandbox is able to directly access the contents of memory
outside of its own memory segments, \eg from the host application or
from another sandbox.
As an example, the following pseudo-assembly program is vulnerable to a breakout attack:

\vspace{1em}
{
  \newcommand\comm[1]{\textnormal{\Small ;\ \ {#1}}}
\begin{tabular}{llll}
   $\jmp\ \textnormal{\uline{end}} \textit{ if } e_{\mathit{check}} $&&$ \comm{if $e_{\mathit{check}}$} $\\
   $\mgen{\rstk}{4} := \rheap $&&$ \comm{\ \ spill $\rheap$ to the stack} $\\
   $\rheap := r_A $&&$ \comm{\ \ and replace its contents with $r_A$} $\\
   $\jmp\ \textnormal{\uline{end}} \textit{ if } \lnot e_{\mathit{check}} $&&$ \comm{else} $\\
   $r_B := \mgen{\rheap}{24} $&&$ \comm{\ \ load a value from the heap} $\\
   \hspace{-0.5em}\uline{end}:
\end{tabular}
}
\vspace{1em}

\noindent
Even though \emph{architecturally} the final load is safe---as the
two conditions are mutually exclusive---\emph{speculatively} we might
(mis)predict and enter both conditional blocks anyway.
Under these conditions, the value in $r_A$ is incorrectly used as the
heap base address, so an attacker that controls the value of $r_A$ can
exploit this behavior to access arbitrary memory---including
memory outside the sandbox.

\mypara{Poisoning attacks}
Even if a sandbox protects its own secrets from leaking architecturally,
it may be speculatively \emph{poisoned} and still leak these secrets on
mispredicted execution paths.
We present the following simple example, where $X$ and $Y$ are arrays of length 64 in the sandbox's
heap and $r_A$ is an index into $X$.

\vspace{1em}
{
  \newcommand\comm[1]{\textnormal{\Small ;\ \ {#1}}}
\begin{tabular}{llll}
   $\jmp\ \textnormal{\uline{end}} \textit{ if } r_A \geq 64 $&&$ \comm{check bound for heap array $X$} $\\
   $r_B := \mgen{\rheap}{X + r_A} $&&$ \comm{\ \ out of bounds if mispredicted} $\\
   $r_C := \mgen{\rheap}{Y + r_B} $&&$ \comm{\ \ leak $r_B$ via memory address} $\\
   \hspace{-0.5em}\uline{end}:
\end{tabular}
}
\vspace{1em}

\noindent
Under architectural execution, any value within $X$ may be leaked due to the final memory access,
but values outside of $X$ are not leaked due to the initial conditional check.
However, during speculative execution, we may incorrectly predict that the branch should fall
through even when $r_A$ is out-of-bounds for $X$. If an attacker is able to control the value
of $r_A$, they can then leak any value in the victim sandbox's heap.

\subsection{Speculative SFI enforcement}

To enforce speculative SFI and prevent breakout and poisoning attacks, Swivel
uses a number of compilation techniques. In addition,
Swivel's SFI system operates on WebAssembly (Wasm) programs;
as such, it can rely on security properties conferred by
well-formed Wasm programs, such as Wasm's memory
regions and its reliance on indirect jump tables.
We describe these properties as well as Swivel's own
techniques for enforcing speculative SFI.

\mypara{WebAssembly guarantees}
A Wasm program's memory at runtime is divided into several distinct
regions: The heap, the stack, and global memory.
The heap in a Wasm program is the only region that can be explicitly ``addressed'';
although Wasm has no pointers, its load and store operations take an \emph{offset}
into the heap region. The base address of the heap is kept in a register.
The Wasm stack is only used for register spills and function return addresses;
it cannot be arbitrarily accessed by a program. All stack spills are known
at compile-time, so all stack accesses are compile-time constant offsets
from the current stack frame, which is kept in a register.
The Wasm global memory holds a program's \emph{jump tables} as well as other
global constants. As such, with the exception of jump table entries, all global
accesses are also compile-time constant offsets into the global region.
Finally, all indirect jumps (or indirect calls) in a Wasm program happen explicitly
via these jump tables: The target is given by an index into
the program's jump table rather than as a direct address.

\mypara{Memory safety}
Swivel employs several techniques, both at compile-time and runtime,
to provide coarse-grained memory safety to sandbox programs
even in the face of speculation.
First, it ``pins'' the heap base register---it prevents this
register from being spilled to the stack and from being
used in general computation---so that the heap base cannot
be corrupted.
Next, it \emph{hardens} heap and jump table loads: Any values used
as offsets into the heap are first truncated to the size
of the heap region before they are added to the heap base---%
offsets into the jump table are similarly truncated.
Other memory accesses, such as for the stack or global constant,
do not need to be hardened, as they only use compile-time constant offsets.
Finally, Swivel's runtime system places \emph{guard pages} around each
of the distinct memory regions, so that any region over- or underflow
(\eg by \emph{return}ing too many times) will immediately halt the program.

\subsection{Speculative CFI and linear blocks}

Swivel offers two distinct approaches for how it enforces speculative CFI.
The first approach, Swivel-SFI, is intended for current x86 processors
and relies heavily on rewriting control flow constructs.
The second approach, Swivel-CET, relies on the Control-flow Enforcement Technology (CET)
extensions developed by Intel in their latest hardware~\cite{shanbhogue2019security}.
Both of these approaches depend on Swivel's concept of \emph{linear blocks}.

\mypara{Linear blocks}
A linear block consists of a sequence of instructions ending in \emph{any} control flow instruction,
and with no other control flow instructions in the block.
When compiling a sandbox program, Swivel first breaks the program into a collection
of such linear blocks.
Swivel then enforces speculative control-flow integrity (CFI) at the granularity
of these linear blocks: It ensures that
all control flow transfers---which always take place at the end of a linear
block---always land at the start of another linear block.

\mypara{Swivel-SFI}
Swivel-SFI's approach provides security, somewhat counterintuitively, by replacing all non-trivial control flow
with indirect jumps. Conditional jumps, for instance, are emulated by selecting
the target block's address based on the relevant condition;
calls and returns are replaced with instructions that save return addresses to
a \emph{separate stack}---distinct from the existing stack memory region and with its own
(pinned) stack pointer register.
Swivel-SFI then protects
speculative control flow by flushing the indirect jump predictor (or \emph{BTB}
for \emph{Branch Target Buffer}~\cite{Canella2019})
upon entering the sandbox.

\mypara{Swivel-CET}
The Swivel-CET implementation makes use of two features from Intel's
CET hardware extensions: The $\cetendbranch$ instruction and the \emph{hardware shadow stack}.
The $\cetendbranch$ instruction provides \emph{forward{\hyp}edge} CFI:
Every control flow instruction (except returns) must land on an \textit{endbranch} instruction,
even when executing speculatively.
For call and return instructions, CET provides a hardware{\hyp}enforced shadow stack:
All calls and returns, in addition to pushing and popping return addresses off the
regular stack, also push and pop return addresses on a separate protected memory region.
When returning, the processor only jumps to a predicted return location if the prediction agrees with
the address popped from the shadow stack~\cite{shanbhogue2019security}.
Finally, Swivel-CET inserts a \emph{register interlock} at every
linear block transition: The register interlock is a sequence of instructions
that detects whether speculative control flow
has been mispredicted, and if so, clears all the memory base registers (\ie the heap base and stack
frame registers).
By doing so, all memory operations following a misprediction are directed to invalid addresses.

\section{Formal model}

To study SFI in the context of speculative execution attacks, we develop a
simple assembly-like language, \emph{\lang}.
We present the syntax of \lang, then formalize its architectural and
speculative semantics.

\subsection{Syntax}
The syntax of \lang{} programs is given in \cref{fig:language:syntax}.
In \lang, expressions are constructed by combining immediate
values $v$ and registers $r$ using basic arithmetic operations
$\oplus$.
\lang supports standard control-flow instructions (conditional and
indirect jumps, function calls and returns), register assignments
($r := e$), memory loads ($r' := \mgen{r}{e}$), and stores
($\mgen{r}{e} := e'$). Memory instructions always access memory at
an offset $e$ from a base register $r$; we mirror Wasm,
which only allows accessing offsets into the distinct memory regions.
To model Swivel implementations,
\lang also supports dedicated instructions $\flush$ (to flush the BTB)
and $\cetendbranch$ (for hardware CFI).

\subsection{Architectural semantics}

We first cover the \emph{architectural semantics} of \lang, which
models the execution of our basic assembly programs \emph{without} any speculative behavior.
The semantics is defined in terms of architectural configurations $\ctx$.
Each configuration $\ctx$ is a quadruple consisting of a program $\P$ mapping values to instructions, a program counter $\pc \in \mathbb{V}$, a register file $\Reg: \mathbb{R} \to \mathbb{V}$ mapping registers to values, and a memory $\Mem: \mathbb{V}\to \mathbb{V}$ that maps memory addresses to values.
We use dot-notation to access a context's elements, \eg $ \ctx.\Mem$ denotes the memory associated with $\ctx$. We use bracket-notation to update an element within a context, \eg $\ctxnext{ \Reg := \Reg' }$ denotes the context obtained by updating the register file to $\Reg'$.
Furthermore, $\ictx{\insn}$ denotes that $\insn$ is the instruction pointed by
the current program counter $\ctx.\pc$; and $ \ctxincr$ denotes the context obtained by
incrementing the program counter of $\ctx$ by 1.

Our architectural semantics is formalized by the $\qstep$ relation in \cref{fig:basic-semantics}, which describes how architectural contexts are modified during the computation.
In the rules, $\eval{e}$ denotes the value of expression $e$ in the context of $\ctx$, and $\rstk$ and $\rheap$ represent the unique \emph{stack pointer} and \emph{heap pointer} registers.
The architectural semantics are straightforward;
for example, to initiate the function call $\call\ i$, rule
\rref{call} decrements the stack pointer
($\vstk = \eval{ \rstk - 1 }$)\footnote{We represent the
  stack growing from higher to lower addresses.}, saves the return address on the stack
($\Mem[\vstk] := \pc + 1$), then jumps to the first
instruction of the function ($\pc := \pc + i$).
Rule \rref{call-indirect} is similar, but for indirect function
calls ($\call\ r$); thus the rule evaluates the address of the function
at run-time ($\vaddr = \eval{ r }$) and directly updates the
program counter ($\pc := \vaddr$).

\begin{figure}[t]
  \centering
  {\small
  \begin{tabular}{rclll}
    \multicolumn{4}{l}{{\bf Basic types}} \\
    \emph{(Values)} & $i,v$ & $\in$ & $\mathbb{V}$  \\
    \emph{(Registers)} & $r$ & $\in$ & $\mathbb{R}$ \\
    \emph{(Operators)} & $\oplus$ & $\in$ & $\bigoplus$\\
    \\
    \multicolumn{4}{l}{{\bf Syntax}} \\
    \emph{(Expressions)} & $e$ & $\in$ & $v \mid r \mid e \oplus e$\\
    \emph{(Instructions)} & $insn$ & $\in$ & $r := e $ & {(assignments)}\\
    & & & $\mid r := \mgen{r}{e}$  & {(memory load)} \\
    & & & $\mid \mgen{r}{e} := e$   & {(memory store)} \\
    & & & $\mid \jmp\ {\pm{i}}$   & {(unconditional jump)} \\
    & & & $\mid \jmp\ {\pm{i}} \textit{ if } e$   & {(conditional jump)} \\
    & & & $\mid \jmp\ r$   & {(indirect jump)} \\
    & & & $\mid \call\ {\pm{i}}$   & {(direct call)} \\
    & & & $\mid \call\ r$   & {(indirect call)} \\
    & & & $\mid \ret$  & {(return)} \\
    & & & $\mid \flush$ & (BTB state flush) \\
    & & & $\mid  \cetendbranch$ & (CET ``endbranch'') \\
  \end{tabular} }
  \caption{Syntax of the \lang language.}%
  \label{fig:language:syntax}
\end{figure}

\begin{figure*}[t]
  {\small
  \begin{mathpar}
    \inferrule[assignment]
    {
      v = \eval{e} \\
    }
    { \ictx{r := e} \qstep \ctxplus{ \Reg[r] := v } }

    \inferrule[load]
    {
      \vaddr = \eval{ \rbase + \eoff } \\
      v = \ctx.\Mem[\vaddr] \\
    }
    { \ictx{r := \m} \qstep
      \ctxplus{ \Reg[r] := v } }

    \inferrule[store]
    {
      \vaddr = \eval{ r_\base + \eoff } \\
      v = \eval{ e } \\
    }
    { \ictx{\m := e} \qstep
      \ctxplus{ \Mem[\vaddr] := v } }

    \inferrule[jump]
    {
    }
    { \ictx{\jmp\ {+i}} \qstep
      \ctxnext{ \pc := \pc + i } }

    \inferrule[jump-cond-taken]
    {
      \eval{ e } \\
    }
    { \ictx{\jmp\ {+i} \textit{ if } e} \qstep
      \ctxnext{ \pc := \pc + i } }

    \inferrule[jump-cond-not-taken]
    {
      \lnot \eval{ e } \\
    }
    { \ictx{\jmp\ {+i} \textit{ if } e} \qstep
      \ctxincr }

    \inferrule[jump-indirect]
    {
      \vaddr = \eval{ r } \\
    }
    { \ictx{\jmp\ r} \qstep
      \ctxnext{ \pc := \vaddr } }

    \inferrule[call]
    {
      \vstk = \eval{ \rstk - 1 } \\
    }
    { \ictx{\call\ i} \qstep
      \ctx \{{
        \begin{tabular}[t]{ll}
          $\Mem[\vstk] := \pc + 1$ &,\\
          $\Reg[\rstk] := \vstk$ &,\\
          $\pc := \pc + i$ &\}\\
        \end{tabular} } }

    \inferrule[call-indirect]
    {
      \vaddr = \eval{ r } \\
      \vstk = \eval{ \rstk - 1 } \\
    }
    { \ictx{\call\ r} \qstep
      \ctx \{{
        \begin{tabular}[t]{ll}
          $\Mem[\vstk] := \pc + 1$ &,\\
          $\Reg[\rstk] := \vstk$ &,\\
          $\pc := \vaddr$ &\}\\
        \end{tabular} } }

    \inferrule[return]
    {
      \vstk = \eval{ \rstk } \\
    }
    { \ictx{\ret} \qstep
      \ctx \{{
        \begin{tabular}[t]{ll}
          $\Reg[\rstk] := \vstk + 1$ &,\\
          $\pc := \Mem[\vstk]$ &\}\\
        \end{tabular} } }
  \end{mathpar}
  }
  \caption{Architectural semantics for {\protect\lang{}}.}%
  \label{fig:basic-semantics}
\end{figure*}

\subsection{Attackers and observations}

To represent the power of attackers to observe and exfiltrate secret data, we have our
semantics emit
\emph{leakage observations} that represent side-channel information an attacker can glean.
The observations emitted by different instructions depends on the \emph{leakage model} we wish to consider.
We consider the following three leakage models, each giving increasing power to an attacker:
\begin{itemize}
  \item \emph{dmem}, where attackers can observe the state of the data cache,
  \item \emph{ct}, where attackers can observe leaks considered by the \emph{constant-time} paradigm~\cite{cauligi2020foundations}, and
  \item \emph{arch}, where attackers can observe all values retrieved from memory~\cite{guarnieri2021contracts}.
\end{itemize}

The \emph{dmem} model is the weakest of the
three models, and considers the data cache as the only viable leakage channel.
In this model, an attacker can observe cache state (specifically, the data cache) using attacks such as \textsc{Prime+Probe}~\cite{tromer2010efficient}, but cannot determine
the control flow trace of a program.
In the \emph{ct} model, we consider an attacker that can observe the standard \emph{constant-time} leakages~\cite{cauligi2020foundations} via timing
or other microarchitectural leaks~\cite{moghimi2019memjam,yarom2017cachebleed,gras2018translation}.
The data cache as well as the control flow trace are visible to the attacker in this model.
Finally, in the \emph{arch} model, we assume the attacker observes all values
loaded from memory. Since the initial memory is the source of all values in the
program, an attacker observing all loaded values during execution is equivalent to an attacker
that sees the complete trace of all values in registers and memory~\cite{guarnieri2021contracts}.

We expose these leakage models via a function $\textsc{Leaks}(\ctx)$
(informally illustrated by \cref{fig:leakage})
that takes as input a configuration $\ictx{\insn}$ and outputs observations
for \emph{each} jump, load, or store operation that occurs during the
semantic execution rule for $\insn$.
For example, the execution of $\ictx{\ret}$ (rule \rref{return} in
\cref{fig:basic-semantics}) contains both a load ($\vstk = \eval{ \rstk }$)
and a jump ($\pc := \Mem[\vstk]$).
Accordingly, under the \emph{ct} model, $\textsc{Leaks}(\ictx{\ret})$ will result in two observations:
$\vstk$, for loading the return address; and $\Mem[\vstk]$, for jumping to that location.

Finally, we include a structure \emph{Obs} in our configuration to collect the sequence of leakage
observations during execution.
We update $\Obs$ with each architectural step using the relation $\qstep_{\mathit{trace}}$ induced by the following rule:
\begin{mathpar}
  \inferrule[trace]
  {
    \ctx \qstep \ctx' \\
    \Obs' = \ctx.\Obs \mathop{\incr} \textsc{Leaks}(\ctx) \\
  }
  { \ictx{\insn} \qstep_{\mathit{trace}} \ctxnext[\ctx']{\Obs'}}
\end{mathpar}
We refer to this extended relation as $\qstep$ for brevity, as it merely adds bookkeeping to the semantics.

\begin{table}[t]
  \centering
  \caption{Informal definition of $\textsc{Leaks}(\ctx)$. The result of
  $\textsc{Leaks}(\ctx)$ for a given leakage model is the aggregate of any
  leakages produced during its execution step.}%
  \label{fig:leakage}
  {\small
  \begin{tabular}{l@{~}l *3l}
    \toprule
    && \multicolumn{3}{c}{Leakage model} \\
    \multicolumn{2}{l}{Effect(s) of $\ictx{\insn} \qstep \ctx'$}
      & \emph{dmem}
      & \emph{ct}
      & \emph{arch} \\
    \midrule
    any jump & ($pc := v$)
        & $\nothing$ & $v$ & $v$ \\
    any load & ($v = \Mem[\vaddr]$)
        & $\vaddr$
        & $\vaddr$
        & $\vaddr, v$ \\
    any store & ($\Mem[\vaddr] := v$)
        & $\vaddr$
        & $\vaddr$
        & $\vaddr$ \\
        \bottomrule
  \end{tabular}
  }
\end{table}

\subsection{Speculative semantics}

\begin{figure*}[t]
  {\small
  \begin{mathpar}
    \inferrule[spec-predict]
    {
      \textsc{IsControlFlow}(\insn) \\
      \ctx \qstep \ctx' \\
      \pc', \ustate' = \textnormal{Oracle}(\ctx) \\
      \mathit{correct} = (\pc' = \ctx'.\pc)
    }
    { \ictx{\insn} \zstep \ctx' \{{
      \begin{tabular}[t]{ll}
        $\pc'$, $\ustate'$, &\\
        $\flag := \ctx.\flag \lor \lnot \mathit{correct}$ &\}\\
      \end{tabular} } }

      \inferrule[spec-step]
      {
        \neg \textsc{IsControlFlow}(\insn) \\
        \ctx \qstep \ctx'
      }
      { \ictx{\insn} \zstep \ctx'  }

      \inferrule[spec-trace]
      {
        \ctx \zstep \ctx' \\
        \Obs' = \ctx.\Obs \mathop{\incr} \textsc{Leaks}(\ctx) \\
      }
      { \ictx{\insn} \zstep_{\mathit{trace}} \ctxnext[\ctx']{\Obs'}}
  \end{mathpar}
  }
  \caption{Speculative semantics for {\protect\lang{}}.}%
  \label{fig:spec-semantics}
\end{figure*}

To reason about speculative leaks, we equip \lang{} with a speculative semantics  that captures the effects of speculatively executed instructions.

We model microarchitectural predictors using a \emph{prediction
  oracle} which abstracts away from the microarchitectural prediction
details.
The oracle is defined in terms of a set of oracle states $\ustate$
(which contains a designated initial state $\bot$) and a
function $\text{Oracle}(\cdot)$ which, given the
current context $\ctx$,
produces the predicted program counter
$\pc'$ and an updated oracle state $\ustate'$.
For simplicity, our speculative semantics does not model
\emph{rollbacks}.
This limited model of speculation is sufficient to analyze
the security guarantees of software and hardware mechanisms for
preventing speculative attacks.

The speculative semantics is formalized by the relation $\zstep$ given in \cref{fig:spec-semantics}.
In the rules defining $\zstep$ configurations, $\Psi$ is extended to store the $\ustate$ of the prediction oracle (which is updated throughout the computation) as well as a simple flag $\flag$ that is set as soon as an oracle prediction is incorrect.
We use this flag in our security proofs to show the
  absence of speculative leaks along mispredicted paths.

The rules in \cref{fig:spec-semantics} demonstrate how
the speculative semantics is obtained from the architectural semantics.
We only consider speculative effects for instructions that modify
control-flow;
otherwise, rule \rref{spec-step} executes instructions
using the architectural semantics.

Rule \rref{spec-predict} describes the speculative execution of
control-flow instructions where the prediction oracle is
invoked to obtain the new program counter $pc'$ and predictor state
$\ustate'$.
In order to detect a misprediction, the rule executes the instruction
using the architectural semantics ($\ctx \qstep \ctx'$), compares
the predicted and the architectural program counters
($\mathit{correct} = (\pc' = \ctx'.\pc)$), then updates the
mispredicted flag accordingly
($\flag := \ctx.\flag \lor \lnot \mathit{correct}$).
Notice that this rule updates the program counter in the final
configuration $\ctx'$ to the value $pc'$ predicted by the oracle: If
this value is incorrect, then the program simply starts executing
instructions along a mispredicted path, as recorded by the flag
$\flag$.

Finally, rule \rref{spec-trace} defines $\zstep_{\mathit{trace}}$ analogously
to $\qstep_{\mathit{trace}}$.
As before, we refer to this extended relation as $\zstep$ for brevity.

\section{Formalizing speculative SFI security}

With the semantics for \lang, we investigate what it means for SFI sandboxing
to be \emph{speculatively secure}. We examine the mitigations implemented by
Swivel in terms of our semantics and we formally define the security properties
that Swivel claims to provide.

\subsection{Speculative SFI security properties}
\label{sec:sfi-attacks}
We present the formal security statements of \lang programs in terms of \emph{non{\hyp}interference properties}.
Generally, a program is \emph{non-interferent} if, for all pairs of initial contexts
that may differ only in sensitive values (\ie any values we don't want to leak
to an attacker), the attacker cannot distinguish the two resulting executions.
Otherwise, the attacker can learn some information about the sensitive values.

Furthermore, we define security with respect to a \emph{class of oracles} $\Omega$.
This allows us to model assumptions about microarchitectural predictors while remaining abstract
over specific predictor implementations:
For example, we can exclude from $\Omega$
any oracles that predict based on memory contents;
or only consider oracles that, for conditional jumps,
will predict one of the two resulting branches.

\mypara{Breakout security}
A breakout attack occurs when an attacker is able to (speculatively or otherwise)
load values from outside its defined memory regions.
In our semantics, we can capture this notion with the \emph{arch} leakage model:
If we consider the sandbox program's own memory regions as benign and all other
memory as sensitive, then a successful breakout attack is equivalent to a sensitive value
appearing in the program's observation trace.

Formally, to prevent breakout attacks, we must show that all sandboxed
programs satisfy non{\hyp}interference under the \emph{arch}
leakage model. We consider two initial
contexts for a program \emph{equivalent} if their respective
memories agree for all sandboxed memory regions; we write this
relation as $\approx_{MR}$.
A program, then, is \emph{breakout secure} (up to $n$ steps) against
a class of oracles $\Omega$ if,
for all $\textnormal{Oracle} \in \Omega$ (and with resulting $\zstep)$,
for all initial contexts $\ctx_1$ and $\ctx_2$:
  \begin{align*}
    \ctx_1 \approx_{MR} \ctx_2 &\textnormal{\ \ and\ \ }
     \ctx_1 \zstep^n \ctx'_1 \\ &\textnormal{\ \ and\ \ }
     \ctx_2 \zstep^n \ctx'_2 \\ &\implies
     \ctx'_1.Obs_{arch} = \ctx'_2.Obs_{arch}
  \end{align*}
where $Obs_{arch}$ is the observation trace given by $\textsc{Leaks}(\cdot)$
for the \emph{arch} model.

\mypara{Poisoning security}
A poisoning attack occurs when a victim sandbox is coerced into
leaking a value during speculative execution that it would not
have leaked architecturally. We capture this notion in our
semantics by comparing the architectural and speculative observation
traces of a program:
Formally, a program is \emph{poisoning secure} (up to $n$ steps) against
a class of oracles $\Omega$ if,
for all $\textnormal{Oracle} \in \Omega$ (and with resulting $\zstep)$,
for all initial contexts $\ctx_1$ and $\ctx_2$:
  \begin{align*}
    \textnormal{If\ \ } & \ctx_1 \qstep \ctx^*_1 \textnormal{\ \ and\ \ } \ctx_2 \qstep \ctx^*_2 \\
    \textnormal{\ \ and\ \ } & \ctx_1 \zstep^n \ctx'_1 \textnormal{\ \ and\ \ } \ctx_2 \zstep^n \ctx'_2 \\
    \textnormal{\ \ and\ \ }& \ctx^*_1.Obs_{ct} = \ctx^*_1.Obs_{ct} \\
     \textnormal{then\ \ } &
     \ctx'_1.Obs_{ct} = \ctx'_2.Obs_{ct} .
  \end{align*}
where $Obs_{ct}$ is the observation trace given by $\textsc{Leaks}(\cdot)$
for the \emph{ct} model.

\subsection{Analyzing Swivel with \lang}

Since Swivel only operates on valid WebAssembly programs, we can make certain
assumptions about the structure of our input programs.
For example, the stack region
(represented in \lang as $\Mem[\rstk + \eoff]$)
is only used
for local variables and register spills; all stack loads and stores use
constant (immediate) offsets from the stack pointer (\ie $\eoff$ for $\rstk$ is always a simple value).
Furthermore,
the heap pointer ($\rheap$ in \lang) is never spilled to the stack, and the
stack pointer ($\rstk$) is only modified when establishing function stack frames.
Finally, since Swivel's runtime surrounds each memory region with guard pages,
we consider any over- or underflow (\eg of the stack) to get stuck.

\begin{figure*}[t]
  {\small
  \begin{mathpar}
    \inferrule[spec-cet-step]
    {
      \lnot \textsc{IsControlFlow}(\insn) \\
      \insn \notin \{\call\ \cdot, \ret\} \\
      \ctx \zstep \ctx' \\
    }
    { \ictx{\insn} \zstep_{\mathit{cet}} \ctx' }

    \inferrule[spec-cet-endbranch]
    {
      \textsc{IsControlFlow}(\insn) \\
      \insn \notin \{\call\ \cdot, \ret\} \\
      \ctx \zstep \ctx' \\
      \ictx[\ctx']{\cetendbranch} \\
    }
    { \ictx{\insn} \zstep_{\mathit{cet}} \ctx' }

    \inferrule[spec-cet-call]
    {
      \ctx \zstep \ctx' \\
      \ictx[\ctx']{\cetendbranch} \\
      \vsstk = \eval{ \rsstk - 1 } \\
    }
    { \ictx{\call\ \cdot} \zstep_{\mathit{cet}}
      \ctx' \{{
        \begin{tabular}[t]{ll}
          $\Mem[\vsstk] := \pc + 1$ &,\\
          $\Reg[\rsstk] := \vsstk$ &\}\\
        \end{tabular} } }

    \inferrule[spec-cet-return]
    {
      \ctx \zstep \ctx' \\
      \vsstk = \eval{ \rsstk } \\
      \ctx'.\pc = \ctx.\Mem[\vsstk] \\
    }
    { \ictx{\ret} \zstep_{\mathit{cet}}
      \ctxnext[\ctx']{ \Reg[\rsstk] := \vsstk + 1 } }
  \end{mathpar}
  }
  \caption{CET semantics for {\protect\lang{}}.}%
  \label{fig:cet-semantics}
\end{figure*}

We analyze both the Swivel-SFI and Swivel-CET implementations on whether or not
they soundly prevent breakout and poisoning attacks.
In general, we want to show that a program, upon leaving any linear block, will always land
on the start of a new linear block. We can use this to inductively extend local block invariants
to cover the whole program.

\subsubsection{Swivel-SFI}
\label{sec:swivel-sfi}
~
\smallskip

Swivel-SFI replaces all (non-trivial) control flow with indirect jumps, flushing
the BTB predictor upon the program's entry.
Since the only relevant predictor in Swivel-SFI is the BTB, we model the $\flush$ instruction by clearing
the entire $\ustate$ to the empty state $\bot$:
  \begin{mathpar}
    \inferrule[spec-flush]
    {
    }
    { \ictx{\flush} \zstep \ctxplus{
      \ustate := \bot } }
  \end{mathpar}
Flushing $\ustate$ will not prevent misprediction: For example, depending on the choice of
$\textnormal{Oracle} \in \Omega$, the prediction oracle may still predict an incorrect
target when $\ustate = \bot$.
It may, however, limit an attacker attempting
to mistrain victim predictors.
After a flush, BTB predictions have no state to rely on beyond
the program itself; Swivel thus assumes that any given jump instruction can only be
trained to historically valid targets
from that $\pc$ location.
For our analysis, we limit $\Omega$ to such oracles.

\mypara{Breakout security}
Swivel-SFI hardens all memory operations; thus if we execute a linear block,
we know the block itself will be secure from breakout attacks.
We thus need only show
that when a Swivel-SFI program exits one linear block, it will always
start at the top of another linear block.

Architecturally, since all (forward-edge) indirect jumps are implemented via hardened jump
table lookups, all such control flow will always target valid linear blocks.
All backward-edge jumps (\ie returns) are implemented via popping and jumping
to an address from the \emph{separate stack}, which itself
is only accessible via the pinned separate stack register. Thus addresses on this separate stack
cannot be otherwise overwritten, and so will always point to valid linear blocks.

Finally, since we limit $\Omega$ to oracles that predict historically
(architecturally) valid targets for any given $pc$, all oracle predictions will
also only target valid linear blocks.

\mypara{Poisoning security}
Unfortunately, even with our selection of $\Omega$,
we cannot soundly prove that Swivel-SFI programs are secure from poisoning attacks.
As a trivial example, consider the program demonstrating a poisoning attack in
\cref{sec:sfi-attacks-background}.
Even after it is converted to use an indirect jump to replace the conditional branch, it may
still mispredict the direction of the condition and execute the vulnerable loads---flushing
the BTB does not prevent mispredictions from happening.
However, by flushing the BTB, Swivel-SFI claims to prevent an attacker from \emph{actively}
mistraining a predictor---\ie an attacker cannot \emph{force} the victim sandbox
to mispredict, and any secret leakage would be purely opportunistic~\cite{swivel}.
Our current framework does not distinguish active attackers in its security
model; we leave formal analysis of active attackers to future work.

\subsubsection{Swivel-CET} \label{sec:swivel-cet}
~
\smallskip

We formalize the CET hardware extensions as an augmented step relation $\zstep_{\mathit{cet}}$
built on top of our prior speculative relation $\zstep$, shown in \cref{fig:cet-semantics}.
The special semantics for CET only affect control-flow instructions (rule \rref{spec-cet-step}).

For all forward-edge control flow,
the CET hardware checks that the instruction at the target address---even
when speculatively predicted---is the special $\cetendbranch$ instruction.
We represent this in rules \rref{spec-cet-endbranch} and \rref{spec-cet-call}
with the clauses $\ctx \zstep \ctx'$ and $\ictx[\ctx']{\cetendbranch}$.

The CET hardware protects backward-edge jumps using a special hardware-managed
\emph{shadow stack}---similar to Swivel-SFI's separate stack---and which
is indexed through an otherwise-inaccessible shadow stack register $\rsstk$.
Furthermore, upon a return, the shadow stack value must agree with the predicted
return value for execution to proceed. We formalize this in
rules \rref{spec-cet-call} and \rref{spec-cet-return}.

\mypara{Breakout security}
As with Swivel-SFI, Swivel-CET masks all memory operations within a linear block.
By placing $\cetendbranch$ instructions only at the tops of linear blocks, and by relying on the CET
shadow stack,
Swivel-CET provides CFI at the linear block level.

\mypara{Poisoning security}
To prevent poisoning attacks, Swivel-CET inserts instructions at the beginning
and end of every linear block to form a \emph{register interlock}, implemented as follows:
Each linear block in the program is given a unique label.
At the end of each block, Swivel-CET inserts instructions to dynamically
calculate and save the label of the target block.
For example, just before a conditional branch, the condition expression is used to select between the
two target block labels.
Then, at the start of each block, Swivel-CET inserts instructions to compare the
stored target label to the label of current block.
If the labels do not match, all memory base registers (\ie $\rheap$ and
$\rstk$) are set to a guard page address $\bot$.
Thus any further (data) memory accesses will be stuck and cannot leak
any values.
With register interlocks in place, we can show that if
$\ctx.\flag$ is set, then all following memory operations cannot leak.

However, while this prevents leaking via \emph{memory} operations, this does not stop
leakages via \emph{control flow}.
For example,
if a sandbox secret is already in a register before we mispredict,
then a later linear block may still branch on this register, leaking the secret value.
Thus we can only prove poisoning security for Swivel-CET with respect to the weaker
\emph{dmem} leakage model instead of the stronger \emph{ct} leakage model.

\section{Conclusion}
We present the first formal framework for SFI security in the face of Spectre attacks.
Our language, \lang, is expressive enough to verify the security claims
of the Swivel sandox system; by formalizing Swivel's security properties,
we reveal which of its security claims it soundly upholds, as well
as the explicit assumptions about hardware execution that Swivel relies on.
We plan to extend and apply our framework to analyze the security claims of other 
sandboxing techniques that claim security against Spectre attacks~\cite{shen2019restricting,Jenkins20,site-isolation}.

\begin{acks}
  We thank Aastha Mehta, Anjo Vahldiek-Oberwagner, and Shravan Narayan for
  their valuable insights and input in the development of \lang.
  This work was supported in part by gifts from Cisco and Intel;
  by the NSF under Grant Numbers CCF-1918573 and CAREER CNS-2048262;
  by the Community of Madrid under the  project S2018/TCS-4339 BLOQUES;
  by the Spanish Ministry of Science, Innovation, and University under the
  project RTI2018-102043-B-I00 SCUM and the Juan de la Cierva-Formaci\'on grant
  FJC2018-036513-I;
  by the German Federal Ministry of Education and Research (BMBF) through
  funding for the CISPA-Stanford Center for Cybersecurity;
  and by the CONIX Research Center, one of six centers in JUMP, a Semiconductor
  Research Corporation (SRC) program sponsored by DARPA.
\end{acks}

\newpage
\bibliography{bib}

\end{document}